\documentclass{emulateapj}

\newcommand\hi{\hbox{H\,{\sc i}~}}

\newcommand\hei{\hbox{He\,{\sc i}~}}
\newcommand\heii{\hbox{He\,{\sc ii}~}}

\def\lsim{~\raise0.3ex\hbox{$<$}\kern-0.75em{\lower0.65ex\hbox{$\sim$}}~}
\def\gsim{~\raise0.3ex\hbox{$>$}\kern-0.75em{\lower0.65ex\hbox{$\sim$}}~}

\def\lbrack2{[\![}
\def\rbrack2{]\!]}

\def\nhi{{N_{\rm HI}}}
\def\xion{{x_{\rm HII}}}
\def\tf{{\Delta t_{\rm f}}}
\def\tr{{\Delta t_{\rm r}}}

\def\vcirc{{v_{\rm circ}}}

\def\nug{{\nu_{\rm g}}}
\def\nugp{{\nu_{{\rm g}+1}}}
\def\alphag{{\alpha_{\rm g}}}
\def\i0g{{I_{0\rm g}}}
\def\e0g{{E_{0\rm g}}}

\def\AA{{A}}

\def\km{{\rm\,km}}
\def\kms{{\rm\,km\,s^{-1}}}
\def\kpc{{\rm\,kpc}}
\def\mpc{{\rm\,Mpc}}
\def\s{{\rm\,s}}
\def\ev{{\rm\,eV}}
\def\msun{{\rm\,M_\odot}}

\def\pc{{\rm\,pc}}
\def\cm{{\rm\,cm}}
\def\erg{{\rm\,erg}}
\def\ster{{\rm\,sr}}
\def\hz{{\rm\,Hz}}
\def\sr{{\rm\,sr}}

\shorttitle{Ionization Structure and Kinematics of DLAs}
\shortauthors{Razoumov et al.}

\begin{document}


\title{Adaptive Mesh Refinement Simulations of the Ionization
  Structure and Kinematics of Damped Ly$\alpha$ Systems with
  Self-consistent Radiative Transfer}


\author{Alexei O. Razoumov\altaffilmark{1,2,3}}
\email{razoumov@phy.ornl.gov}

\author{Michael L. Norman\altaffilmark{4}}
\email{mnorman@cosmos.ucsd.edu}

\author{Jason X. Prochaska\altaffilmark{5}}
\email{xavier@ucolick.org}

\author{Arthur M. Wolfe\altaffilmark{4}}
\email{awolfe@ucsd.edu}


\altaffiltext{1}{Physics Division, Oak Ridge National Laboratory, Oak
  Ridge, TN 37831-6354}
\altaffiltext{2}{Department of Physics and Astronomy, University of Tennessee,
  Knoxville, TN 37996-1200}
\altaffiltext{3}{Joint Institute for Heavy Ion Research, Oak Ridge National
  Laboratory, Oak Ridge, TN 37831-6374}
\altaffiltext{4}{Center for Astrophysics and Space Sciences, University of
  California, San Diego, CA 92093}
\altaffiltext{5}{UCO/Lick Observatory, UC Santa Cruz, Santa Cruz, CA 95064}


\begin{abstract}
We use high resolution Eulerian hydrodynamics simulations to study
kinematic properties of the low ionization species in damped
Ly$\alpha$ systems at redshift $z=3$. Our adaptive mesh refinement
simulations include most key ingredients relevant for modeling neutral
gas in high-column density absorbers: hydrodynamics, gravitational
collapse, continuum radiative transfer above the hydrogen Lyman limit
and gas chemistry, but no star formation. We model high-resolution
Keck spectra with unsaturated low ion transitions in two Si II lines
($1526$ and $1808\AA$), and compare simulated line profiles to the
data from the SDSS DLA survey.

We find that with increasing grid resolution the models show a trend
in convergence towards the observed distribution of HI column
densities. While in our highest resolution model we recover the
cumulative number of DLAs per unit absorption distance, none of our
models predicts DLA velocity widths as high as indicated by the data,
suggesting that feedback from star formation might be important. At
$z=3$ a non-negligible fraction of DLAs with column densities below
$10^{21}\cm^{-2}$ is caused by filamentary structures in more massive
halo environments. Lower column density absorbers with $N_{\rm HI} <
10^{21.4}\cm^{-2}$ are sensitive to changes in the UV background
resulting in a $10\%$ reduction of the cumulative number of DLAs for
twice the quasar background relative to the fiducial value, and nearly
a $40\%$ reduction for four times the quasar background. We find that
the mass cut-off below which a large fraction of dwarf galaxies cannot
retain gas after reionization is $\sim 7\times 10^7\msun$, lower than
the previous estimates. Finally, we show that models with
self-shielding commonly used in the literature produce significantly
lower DLA velocity widths than the full radiative transfer runs which
essentially render these self-shielded models obsolete.
\end{abstract}





\section{Introduction}

Damped Ly$\alpha$ absorbers (DLAs) are an excellent tool for probing
structure formation in the early Universe. By definition DLAs are
systems with neutral hydrogen column densities above $2\times
10^{20}\cm^{-2}$, and for many years they have been linked to forming
protogalaxies at high redshifts. With high resolution spectrography,
DLAs can shed light on physical processes and substructure within
individual young galaxies. However, the exact nature of host absorbers
has been debated for many years.

From metal-line kinematic analysis of absorption line widths and
profile asymmetries \citet{prochaska.97} concluded that the observed
DLA population can be best explained with a model in which absorption
is caused by rapidly rotating ($\vcirc\sim 225\kms$) cold thick
($h\sim 0.3R$) disks which have already assembled at high redshifts,
and ruled out the competing models in which DLAs are small ($\vcirc\le
100\kms$) protogalaxies in the process of hierarchical
merger. However, in most models of hierarchical structure formation it
is extremely difficult to have an abundant population of such massive
disks at $z\sim 3-4$.

On the other hand, in the high-resolution (1 kpc spatial and $5\times
10^6\msun$ mass resolution) SPH simulations of a small number of
high-redshift galaxies \citet{haehnelt..98} showed that irregular
protogalactic clumps with $\vcirc\sim 100\kms$ can reproduce the
observed velocity width distribution and absorption profile
asymmetries very well, without the need to invoke rapidly rotating
massive disks. In their models large velocity widths are caused by a
mixture of rotation, random motions, infall, and merging, and the
resulting picture is fully consistent with most standard hierarchical
structure formation scenarios. Despite having achieved very high
numerical resolution, this simulation features a small volume size (2
Mpc) and may not be representative of a typical DLA absorber in the
cosmological context. Another shortcoming of \citet{haehnelt..98}
paper is that their models do not include radiative transfer of the
ultraviolet background (UVB) instead assuming simple self-shielding
above a critical density, an assumption that directly affects the
cross-section of DLAs. In our paper we are going to address this issue
in detail.

Simulations with larger volume sizes inherently suffer from inability
to resolve the lower-mass part of the DLA
distribution. \citet{katz...96} studied the formation of DLAs in a
22.22 Mpc (comoving) volume with SPH simulations with dark matter mass
resolution $2.8\times 10^9\msun$ effectively meaning that only halos
with masses above $\sim 10^{11}\msun$ could be resolved.

\citet{gardner...97a} developed a method to account for absorption in
halos below the numerical resolution of simulations. They used
\citet{press.74} formalism to calculate the number density of
unresolved halos and then convolved this distribution with the
relation between the neutral hydrogen cross-section and halo circular
velocity observed in their SPH simulation to obtain the total
abundance of DLAs in a large simulation volume. We will return to the
important question of DLA cross-sectional dependence on the dark
matter (DM) halo velocity dispersion further in this
introduction. \citet{gardner...97a} find that accounting for
unresolved halos with this resolution correction increases the DLA
incident rate per unit redshift by about a factor of two bringing it
closer to the observed value but still underpredicting the amount of
Lyman limit absorption by approximately a factor of 3.

\citet{gardner...01} used newer SPH simulations and an improved
technique to compute the contribution from unresolved halos and found
that their DLA incidence matches observations as long as they adopt a
$\vcirc\sim 60\kms$ circular velocity cutoff (corresponding to virial
mass $\sim 2\times 10^{10}\msun$ at $z=3$) below which halos cannot
host DLA absorption systems, the value which is somewhat higher than
the more traditional threshold $\vcirc\sim 40\kms$ for suppressing the
formation of dwarf galaxies by a UV photoionization background
\citep{quinn..96,thoul.96}. The highest dark matter mass resolution in
these simulations is $8.3\times 10^8\msun$ immediately suggesting that
halos with masses below ${\rm few}\times 10^{10}\msun$ are not
resolved which is very similar to the suggested cutoff.


On the other hand, \citet{maller...01} used semi-analytic models of
galaxy formation to further investigate the hypothesis that multiple
galactic clumps give rise to the observed kinematics and column
density distributions. They studied several variants of the
exponential disk model in which the sizes of individual galaxies are
based on angular momentum conservation, noting that the resulting
gaseous disks by themselves are too small to produce the observed
kinematics and the overall number density of DLAs. However, with a
simple toy model they showed that a larger covering factor of the cold
gas in these clumps can successfully reproduce the observed properties
suggesting that lines of sight most likely go through tidal tails
caused by mergers.

Furthermore, \citet{prochaska.01} showed that the DLA cross-sections
obtained from the SPH simulations \citep{gardner...01} are not
consistent with the observed velocity width $\Delta v$ distribution,
in fact leading to too many low $\Delta v$ systems. On one hand one
would need a sufficient number of these low-mass systems to match the
total observed number density of DLAs, on the other hand low-mass
systems result in very small $\Delta v$. \citet{prochaska.01}
suggested two possible solutions. One possibility is that the
cross-sectional dependence on the halo mass does not hold below the
resolution limit of \citet{gardner...01} where processes on sub-kpc
scales ranging from various feedback mechanisms to tidal stripping
would disrupt this cross-sectional dependence. The second solution is
that the same feedback mechanisms in lower-mass systems could easily
double the typical velocity widths and are therefore central to
understanding the overall kinematics of DLAs.

These concerns were the focus of the recent simulations by
\citet{nagamine..04a}. Arguing that the relation between the
absorption cross-section and the DM halo mass might not follow the
same trend below $10^{10}\msun$, they performed a series of
simulations covering a wide spectrum of box sizes and mass
resolutions, employing a new conservative entropy formulation of SPH
and a two-phase sub-resolution model for the interstellar medium
(ISM), and varying the amount of feedback from star formation (SF)
with a new phenomenological model for galactic winds. Their highest
resolution run features $216^3$ dark matter particles (and the same
number of gas particles) in a $3.375\mpc$ box giving $2.75\times
10^5\msun$ mass resolution.  whereas their largest simulation volume
is 100 Mpc on a side with $324^3$ dark matter particles and $\sim
10^4$ times lower mass resolution. With high resolution runs in small
boxes, they see DLAs in halos with masses down to $M_{\rm tot}\sim
10^{8.3}\msun$. Below this mass at $z=4.5$ to $z=3$ they notice a
sharp drop-off in the DLA cross-section which they associate with
photoevaporation of gas in these halos by the ionizing UVB and/or to
supernovae feedback. They use the same approach as
\citet{gardner...01} to find the cumulative abundance of DLAs at each
redshift. They fit a functional form to the relation between the DLA
cross-section and the total halo mass observed in their simulations,
and then convolve it with the \citet{sheth.99} parameterization of the
dark matter halo mass function to obtain the total number of DLAs per
unit redshift as a function of halo mass. They find a steeper slope
for the DLA cross-section dependence on halo mass than
\citet{gardner...01} resulting in fewer DLAs from low-mass halos, and
they report a good agreement between the simulated and observed number
of DLAs at $z\ge 3$.

However, none of the simulations mentioned above addressed both the
kinematics and the gas cross-section properties of DLAs in the same
numerical output. We will consider these issues in our current
paper. We also compute self-shielding of protogalaxies from the UVB
with a new radiative transfer algorithm.

Before we proceed to describe our models, it is useful to review how
the numerical simulations mentioned above accounted for shielding from
the UVB beyond the hydrogen Lyman edge. \citet{haehnelt..98} have
adopted a simple scheme to mimic the effect of self-shielding based on
correlation between the column density and the physical density
predicted by earlier numerical simulations in the optically thin
regime. Arguing that self-shielding becomes important above an \hi
column density of $10^{17}\cm^{-2}$, which corresponds to the
absorption-weighted density $10^{-3}-10^{-2}\cm^{-3}$ along the line
of sight, they assume that all gas above a density threshold of
$10^{-2}\cm^{-3}$ is self-shielded and fully neutral, and lower
density gas is ionized.

\citet{katz...96} developed a more complex self-shielding
correction. During actual simulation they compute the species
abundances from ionization equilibrium in the gas that is optically
thin at the Lyman limit everywhere in the volume. Then, at the
post-processing stage they correct projected \hi maps pixel by pixel
recomputing ionizational equilibrium with proper attenuation of the
UVB in a plane-parallel slab with the same total hydrogen column
density and the same physical size, and therefore accounting for
shielding in optically thick systems. According to \citet{katz...96},
at column densities $10^{16}-10^{17}\cm^{-2}$ the correction is small
and varies between $1\%$ and $10\%$, whereas at higher column
densities, $N_{\rm HI} >> 10^{17}\cm^{-2}$, this correction can be as
large as a factor of 100. \citet{gardner...97a,gardner...01} all use
this self-shielding correction analyzing models initially computed
with ionizational equilibrium in an optically thin gas.

On the other hand, \citet{nagamine..04a} do not apply any
self-shielding correction at all arguing that DLAs with column
densities above $\sim 10^{20}\cm^{-2}$ are fully neutral even without
the correction -- an assumption that, in our view, may not hold in all
situations, e.g., for dense shocks in tidal tails which can
nevertheless be subjected to the photoionizing background. Instead,
they use a more complicated two-phase ISM model, in which they compute
the neutral hydrogen fraction with the standard optically thin
approximation and a uniform UVB if the local density is below a
density threshold $\rho_{\rm th}$ which marks the onset of cold cloud
formation. Above $\rho_{\rm th}$ they use a two-phase model in which
the \hi fraction is a function of the local cooling and SF rates and a
few parameters including feedback from supernovae.

In this paper we present new high resolution simulations of damped
Ly$\alpha$ systems at $z=3$. To resolve individual galaxies, we use a
customized version of the adaptive mesh refinement (AMR) Eulerian
hydrodynamics cosmology code \emph{Enzo}
\citep{bryan.99,oshea......04} which includes simultaneous radiative
transfer of the UVB beyond the hydrogen Lyman edge on all levels of
grid refinement. Our goal is to see whether very high spatial
resolution models with sophisticated physics can reproduce the
observed kinematic and statistical properties of the low ionization
DLA metal lines.

We do not include SF feedback in the present study, instead focusing
on the physical complications associated with radiative transfer in
galaxy formation models with limited numerical resolution. We are
planning to include feedback into our future models.

Unlike earlier DLA simulations \citep{nagamine..04a, gardner...01,
haehnelt..98} which considered only absorption by neutral gas in
isolated (and mostly virialized) galaxies, we do not rule out the
hypothesis that at high redshifts $z\ge 3$ a significant fraction of
DLA absorption is caused by \hi outside the virial radii of the
galaxies. There is evidence of active galaxy assembly at those
redshifts, and while DLAs are most likely associated with the neutral
gas from which galaxies form, there is a certain possibility that DLA
absorption is caused by \hi in tidal tails in mergers
\citep{maller...01}, or even in filaments along which the gas is still
falling into dark matter potential wells. \citet{haehnelt..98}
addressed the possibility that DLAs are not necessarily caused by
virialized systems, but they assumed a one-to-one correspondence
between the local baryon density and the ionizational state of
hydrogen which automatically placed all neutral gas into small
self-shielded protogalactic clumps. Similarly, \citet{gardner...01}
and \citet{nagamine..04a} identified all DM halos in their simulations
and used a relation between the DLA cross-section and the total mass
of associated halos to compute the number of DLAs per unit redshift.

Since our models include full angular-dependent radiative transfer, we
expect the association between the distribution of neutral hydrogen
and dark matter halos to emerge as we move to lower redshifts and
higher levels of grid refinement, and consequently as more
intergalactic HI either settles down in halos or gets destroyed by the
UVB, but we do not put this association into our models in any way. In
fact, the existence of high-velocity clouds in the halo of our own
Milky Way Galaxy at present day (e.g. Blitz et al. 1999 and references
therein) implies that we can expect to see extended neutral gas
configurations at high redshifts.

This paper is organized as follows. In section~\ref{Simulations} we
describe our numerical simulations, along with the two-stage radiative
transfer algorithm, the choice of the UVB, and our spectrum generation
routine. Sections~\ref{Results} and ~\ref{Summary} present our
results and conclusions.

\section{Simulations}\label{Simulations}

All simulations were run for the flat $\Lambda$CDM cosmology with
$\Omega_{\rm m}=0.3$, $\Omega_{\rm\Lambda}=0.7$, $\Omega_{\rm
b}=0.045$, $h=0.67$, for a primordial power spectrum $\sigma_8=0.9$
and $n_{\rm s}=1$. All models are summarized in Table 1 which lists
the simulation volume size, the base grid size, the maximum number of
levels of refinement, the total number of grids at the end of each run
at $z=3$, grid resolution, and mass resolution of each model which is
simply the mass of a DM particle. Most of the models include full
radiative transfer, except for runs C2s and C2t which were computed
with self-shielding above density $10^{-2}\cm^{-3}$ and in the
optically thin regime, respectively.

\begin{table}
  \begin{center}
    \caption{List of models.\label{table1}}
    \begin{tabular}{ccccccc}
      \tableline\tableline
      Model & $L$, $h^{-1}{\rm Mpc}$ & Base grid\tablenotemark{a} &
      $N_{\rm levels}$ &
      $N_{\rm grids}$ & $\Delta x$, $h^{-1}{\rm kpc}$ & $\Delta m$, $M_\odot$\\
      & comoving &&& at $z=3$ & comoving &\\
      \tableline
      \multicolumn{7}{c}{2 $h^{-1}{\rm Mpc}$ volume}\\
      \tableline
      A1 & 2 & $128^3$ & 6 & 13,160 & 0.25 & $4.0\times 10^5$\\
      \tableline
      \multicolumn{7}{c}{4 $h^{-1}{\rm Mpc}$ volume}\\
      \tableline
      B1 & 4 & $128^3$ & 6 & 14,150 & 0.5 & $3.2\times 10^6$\\
      \tableline
      \multicolumn{7}{c}{8 $h^{-1}{\rm Mpc}$ volume}\\
      \tableline
      C-1 & 8 & $128^3$ & 4 & 13,228 & 4 & $2.6\times 10^7$\\
      C0 & 8 & $128^3$ & 5 & 14,151 & 2 & $2.6\times 10^7$\\
      C1\tablenotemark{b} & 8 & $128^3$ & 6 & 14,865 & 1 & $2.6\times 10^7$\\
      C2\tablenotemark{c}  & 8 & $128^3$ & 7 & 15,735 & 0.5 & $2.6\times 10^7$\\
\tableline
\end{tabular}
\tablenotetext{a}{Number of base grid cells is always the same as the total
  number of dark matter particles}
\tablenotetext{b}{Three different UVBs were used with this model: one
  with the full stellar ($\beta_*=1$) and full quasar ($\beta_{\rm
    q}=1$) components from Fig.~\ref{fig:uvb02} (C1), one with
  $\beta_*=1$ and $\beta_{\rm q}=2$ (C1b), and finally one with
  $\beta_*=1$ and $\beta_{\rm q}=4$ (C1c).}
\tablenotetext{c}{In addition to the run with full UVB transfer (C2),
  two other models were computed: one with complete self-shielding
  above $10^{-2}\cm^{-3}$ (C2s), and one with the optically thin
  approximation throughout the entire simulation volume (C2t).}
\end{center}
\end{table}

We identify the following three parameters in our calculations: grid
resolution, box size/mass resolution, and the UVB amplitude. We
consider the box sizes of $2h^{-1}$, $4h^{-1}$ and $8h^{-1}$ Mpc, with
base grid resolution $128^3$ and up to seven levels of refinement by a
factor of two. In {\emph Enzo} the total number of DM particles is
always the same as the number of base grid cells $N_{\rm base}^3$
resulting in mass resolution

\begin{equation}
\Delta M\approx 3.2\times 10^6
\left(L_{\rm box}\over\mpc\right)^3
\left(N_{\rm base}\over 32\right)^{-3}\msun.
\end{equation}

\noindent
Therefore, for a fixed base grid, mass resolution is determined by the
current box size.


\subsection{Two-stage radiative transfer}\label{RT}

We use a two-step algorithm for radiative transfer. First, we evolve
all simulation volumes solving the radiative transfer equation in a
low angular resolution mode simultaneously with the equations of
hydrodynamics. To correct for low resolution, we apply a high angular
resolution transfer filter iteratively to our solutions at $z=3$ to
improve on the equilibrium position of HI, HeI and HeII ionization
fronts.

We developed a UVB radiative transfer module for the parallel AMR
cosmology N-body Eulerian hydrodynamics code \emph{Enzo}. The fluid
flow equations are solved with the PPM scheme \citep{colella.84} on a
comoving grid (see for \citet{oshea......04} for details), and
chemical abundances for various ionization and molecular states of
hydrogen and helium are solved with a chemical reaction network with 9
species and 28 reactions \citep{anninos...97, abel...97}. Each level
of grid refinement has twice the spatial resolution of the previous
level. We solve the radiative transfer equation with a photon
conserving scheme self-consistently at each level of resolution
carrying fluxes explicitly from parent grids to all of its
subgrids. On each grid, radiative transfer is computed with a timestep
which is usually significantly smaller than the hydro timestep and is
adjusted adaptively to obtain the best balance between accuracy and
the speed of calculations. Due to complexity of the setup we have not
included point source radiation in our models yet, and also we limit
transport of background radiation to simple sweeps along the
xyz-coordinate axis. We then post-process \emph{Enzo} output with much
higher angular resolution transfer. Below we describe these two steps
in detail.



\subsection{Stage one: time stepping and parallelization of the base grid}

By definition, radiation propagates at the speed of light. In an
explicit advection numerical scheme the Courant condition would
necessarily require prohibitively small timesteps to guarantee
stability. However, our photon conservation technique is inherently
stable, and from this standpoint there is no need to take very small
timesteps, but accuracy of the solution is an entirely different
issue. In many cases where the supply of ionizing photons greatly
exceeds the recombination rate I-fronts can propagate at or close to
the speed of light, and even in this regime our photon conservation
will give an accurate solution as long as the radiation-chemistry
timestep is small enough. How do we ensure that we provide timesteps
small enough to balance accuracy and speed of calculations?

Let us first describe our numerical setup. Since in general radiation
driven fronts can propagate much faster than the fluid flow, radiative
transfer and chemistry normally have to be iterated multiple times per
hydro time step. On each subgrid, independently of the resolution
level, first we perform a hydrodynamical update at a fixed timestep
$\tf$ determined from the hydrodynamical Courant condition on that
subgrid. We then proceed to compute photon conservation in all
directions on a much smaller timestep $\tr=\tf/(2^n-1)$ where $n$ is
some positive integer number. We then update temperature and solve the
chemical rate equations on the same small $\tr$. Next we compare
distributions of some state variable, e.g., a fraction $\xion$ of
ionized hydrogen on our subgrid. If the maximum change in $\xion$
during $\tr$ on the subgrid does not exceed $10\%$ then we increase
$\tr$ by a factor of two, otherwise keep the same small $\tr$, and do
another radiation-chemistry update, and repeat the whole procedure of
these updates until we reach the end of the hydro step $\tf$. Ideally,
if there are no I-fronts on the subgrid we can do the entire
radiation-chemistry calculation with $n$ subcycles for each hydro
step. On the other hand, in the worst case scenario with fast I-fronts
we will need $(2^n-1)$ subcycles. The exact number of subcycles
between $n$ and $(2^n-1)$ is determined automatically by the code
based on maximum relative changes in $\xion$ in our subvolume.

To determine a suitable value of $n$, we ran a number of tests with
$\tr=\tf/(2^n-1)$ fixed throughout the entire run from high to
moderate redshifts. We observe fast convergence as the number of
subcycles reaches few tens, but the solution is accurate enough even
for $(2^n-1)\sim 10$. In all our calculations we use $n=4$ varying the
number of subcycles between $n$ and $(2^n-1)$ as described above.



\begin{figure}
\epsscale{1.}
\plotone{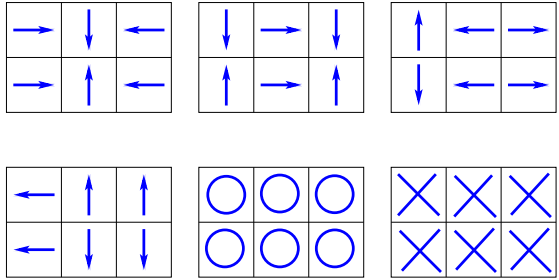}
\caption{A sequence of directional updates for the $3\times 2\times 1$
  volume decomposition.\label{fig:directions}}
\end{figure}

Ionizing photons in our model all originate at the edge of the base
grid, and propagate inward. Parallelization in \emph{Enzo} is done
through volume decomposition in which the lowest resolution base grid
is subdivided equally between all processors. Since we treat radiation
explicitly, solutions to the radiation field in all directions on all
processors have to be connected to each other at every
radiation-chemistry subcycle, and at the same time we need to minimize
the amount of idle time each processor spends waiting for flux updates
from neighboring volumes. To achieve this goal of optimal
parallelization and best load balancing while simultaneously
transporting photons in all directions, we constructed the following
algorithm. At the beginning of every radiation-chemistry subcycle each
processor scans all angular directions in some preset order checking
if input fluxes (from the boundary or from another processor) are
available in that direction and if radiative update has not been done
yet. When it finds such an angle, it performs radiative transfer in
that direction, while all other processors are doing their own
updates. If no input flux in any direction is available then the
processor sits idle for the duration of this directional
subcycle. When the subcycle is done, fluxes are passed to neighboring
volumes, and a new subcycle begins. To further illustrate this idea,
we drew a sequence of directional updates for the $3\times 2\times 1$
volume decomposition in Fig.~\ref{fig:directions}. In this particular
example we have a $100\%$ parallelization efficiency, which is clearly
not always achievable for larger decompositions.

To summarize, in the hierarchy of events on a base grid every single
hydrodynamical step is followed by a series of radiation-chemistry
updates each of which in turn contains directional subcycling with
flux exchanges among individual processors. When a spatial region is
refined, radiation fluxes for both stellar and quasar components in
all six directions are passed from parent grids to subgrids, along
with all hydrodynamical variables.





\subsection{Stage two: high angular resolution transfer}

Our radiation hydrodynamics (RHD) calculation is parallelized via
domain decomposition, in which we pass radiative intensities from one
processor to another on the base grid, and in turn on each processor
we advect these intensities along the AMR grid hierarchy. This
approach results in a fairly complex algorithm, forcing us to limit
RHD transport only to the six directions along the major axes. The
total number of photons entering the volume in plane-parallel fluxes
is kept the same as if we had an isotropic incident distribution,
therefore those photons which burn their way into a denser environment
at small solid angles tend to over-heat and over-ionize the medium. At
the same time this approximation naturally leads to formation of
shadows behind self-shielded regions which in reality could be exposed
to ionizing radiation. To compensate for this shortcoming, we
post-process our 3D ionization and temperature distributions with a
much higher angular resolution radiative transfer code based on a new
fully threaded transport engine algorithm (FTTE) by
\citet{razoumov.05}.

At each output redshift we convert the block-structured AMR output of
our coupled RHD runs to a fully threaded format and use it as a first
guess to compute radiative transfer with FTTE along 192 directions
chosen to subdivide the entire sphere into equal solid angle
elements. We then use this updated radiation field to compute the
ionizational structure and temperature, and then iterate in radiative
transfer and chemistry until we find the equilibrium positions of HI,
HeI and HeII ionization fronts at a given redshift. We find that $\sim
20$ iterations are needed for convergence.

\subsection{Adopted UVB and frequency dependence}

Constraints on the UVB at high redshifts come from the Ly$\alpha$
forest studies. \cite{scott...00} used the proximity effect in quasar
absorption spectra to derive the UVB amplitude assuming that it has
the same spectrum as individual quasars. Using Ly-alpha emission line
redshifts, they get the value $J_\nu=1.4^{+1.1}_{-0.5}\times
10^{-21}\erg\s^{-1}\cm^{-2}\hz^{-1}\sr^{-1}$ for $1.7<z<3.8$. On the
other hand, with OIII and MgII redshifts they get a lower value
$J_\nu=7.0^{+3.4}_{-4.4}\times
10^{-22}\erg\s^{-1}\cm^{-2}\hz^{-1}\sr^{-1}$ for $1.7<z<3.8$,
corresponding to HI photoionization rate $1.9^{+1.2}_{-1.0}\times
10^{-12} \s^{-1}$.

More recently, \citet{tytler..........04} and
\citet{jena..............04} build a concordance model of the
Ly$\alpha$ forest at $z=1.95$ using a series of hydrodynamical
simulations with grid sizes up to $1024^3$ and box sizes up to 76.8
Mpc to reproduce the observed flux decrement from the low-density
intergalactic medium (IGM) alone. Their UVB corresponds to an
ionization rate per \hi atom of $\Gamma_{912}=(1.44\pm 0.11)\times
10^{-12} \s^{-1}$, which is slightly higher than the earlier
estimates, e.g., the prediction by \citet{madau..99} with 61\% from
QSOs and 39\% from stars. The redshift evolution of this concordance
model can be found in \citet{paschos.04}. A more recent study by
\citet{kirkman..........05} extends this higher estimate of
$\Gamma_{912}\sim 1.4\times 10^{-12} \s^{-1}$ to the redshift range
$2.2<z<3.2$ translating into $J_\nu\sim 5\times
10^{-22}\erg\s^{-1}\cm^{-2}\hz^{-1}\sr^{-1}$.

\begin{figure}
\epsscale{1.1}
\plotone{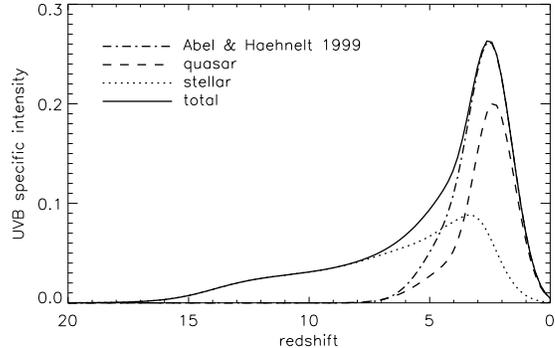}
\caption{Amplitude of the assumed fiducial UVB at $13.6\ev$ in
  standard units of $10^{-21}
  \erg\s^{-1}\cm^{-2}\ster^{-1}\hz^{-1}$. Shown are the quasar
  (dashed line) component $I_{21,{\rm q}}$ and the stellar (dotted
  line) component $I_{21,*}$, and their sum (solid line) at the
  hydrogen Lyman edge for $\beta_*=\beta_{\rm q}=1$. For comparison
  we also plotted the total UVB from \citet{abel.99} (dash-dotted
  line).\label{fig:uvb02}}
\end{figure}

The background in the vicinity of DLAs in a cluster that does not host
a quasar is most likely going to be smaller than these average values,
due to the geometrical dilution factor, and local attenuation in the
cluster. In our other models not listed in Table 1 we experimented
with a variety of UVB spectra and amplitudes, and as a reference we
decided to use a smaller background based on \citet{abel.99}. We
included both stellar and quasar ionizing photons adopting a
two-component power-law UVB



\begin{equation}
I_\nu=
\beta_*
I_{21,*}\left(\frac{\nu}{\nu_L}\right)^{-\alpha_*}+
\beta_{\rm q}
I_{21,{\rm q}}\left(\frac{\nu}{\nu_L}\right)^{-\alpha_{\rm q}}
\end{equation}

\noindent
where $\alpha_*=5$, $\alpha_{\rm q}=1.8$, and $\beta_*$ and
$\beta_{\rm q}$ are the model parameters. We extend the stellar and
quasar components of \citet{abel.99} to higher redshifts with the
expressions

\begin{eqnarray}
I_{21,{\rm q}} &=& Q + a (S-Q),\\ \nonumber
I_{21,*} &=& \left(S+a(Q-S)\right)\left(1-\tanh(0.5z-7)\right)/2,
\end{eqnarray}

\noindent
where

\begin{eqnarray}
Q &=& {10 \exp[-(z/2.5)^3]\over 1+(7/(1+z))^4},\\ \nonumber
S &=& {1-b\over 2}{\exp[-(z/4)^3]\over 1+(7/(1+z))^4}+\\ \nonumber
&& {b~(1+z)^{3.35}\over 106.4}\exp\left[-(z-0.5)^2\over 1+(z+2.09)^{2.075}/16\right],\\ \nonumber
a &=& 0.3 \exp[-(z-4.5)^2/4],\\ \nonumber
b &=& 1+\tanh(1.5z-6.3).
\end{eqnarray}

\noindent
We plotted $I_{21,*}$, $I_{21,{\rm q}}$ and the background from
\citet{abel.99} in Fig.~\ref{fig:uvb02}. We allow for a substantial
stellar component at $z>8$, which is consistent with the current
theoretical views of the global history of SF in a $\Lambda$CDM
Universe \citep{springel.03} in light of the discovery of a large
optical depth to electron scattering by WMAP \citep{kogut03}
suggesting presence of bright ionizing sources at $z\sim 15-17$
\citep{wyithe.03,haiman.03,hui.03,ciardi..03}. As the main focus of
this paper is to study the resolution effects of radiative transfer in
galaxy formation models, we do not explore other possible star
formation histories, but undoubtedly the cumulative history of star
formation at $z>6$ might have an important effect on the observable
properties of galaxies even at lower redshifts.

We use three frequency bands, $13.6-24.6\ev$, $24.6-54.4\ev$, and
above $54.4\ev$, to compute \hi, \hei and \heii ionization. Inside
each frequency band $[\nug,\nugp]$ the transport variable is the
intensity at frequency $\nug$ weighted by a corresponding
power-index-dependent factor

\begin{equation}
f_{\rm g}={1-\left(\nugp/\nug\right)^{1-\alphag}\over\alphag-1},
\end{equation}

\noindent
where ${\rm g}=1,2,3$. All radiation-related rates (photoionization,
photoheating, and absorption) also depend on $\alphag$. Finally,
inside each frequency band we use a single transport variable
describing both stellar and quasar components with an effective index
$\alphag$ constructed to ensure that the total energy inside the band
is equal to the sum of the two individual components.


\subsection{Spectrum generation}\label{sec:spectrum}

To analyze the statistical properties of DLAs, we drew 100,000 random
(through a random point in a random direction) lines of sight through
the entire grid hierarchy of each model using the highest resolution
cells available at each point along the line, and studied all
absorption systems with HI column densities $\nhi\ge 2\times
10^{20}\cm^{-2}$. For analysis we use two ``low ion'' lines of Si II
at $1526\AA$ and $1808\AA$. Since the latter line has a significantly
lower oscillator strength, we use it only in absorbers with
$\log(\nhi/\cm^{-2})>20.6$, and for the lower column densities use the
former transition. We assume a uniform abundance throughout the volume
with metallicity $[{\rm Si/H}]=-1.3$ \citep{wolfe..05}. Noise is added
to our artifical spectra such that the resultant signal-to-noise ratio
of each $1\kms$ pixel is 20:1. We define the Si II line width as the
width of the central part of the profile responsible for $90\%$ of the
integrated optical depth as described in \citet{prochaska.97}. In the
unlikely case that a line of sight crosses multiple absorbers, we
consider two components to be caused by a single DLA if the separation
between the centers of the two corresponding line profiles is less
than $400\kms$, otherwise we just analyze the strongest component.

\section{Results}\label{Results}

\begin{figure*}
\epsscale{1.}
\plotone{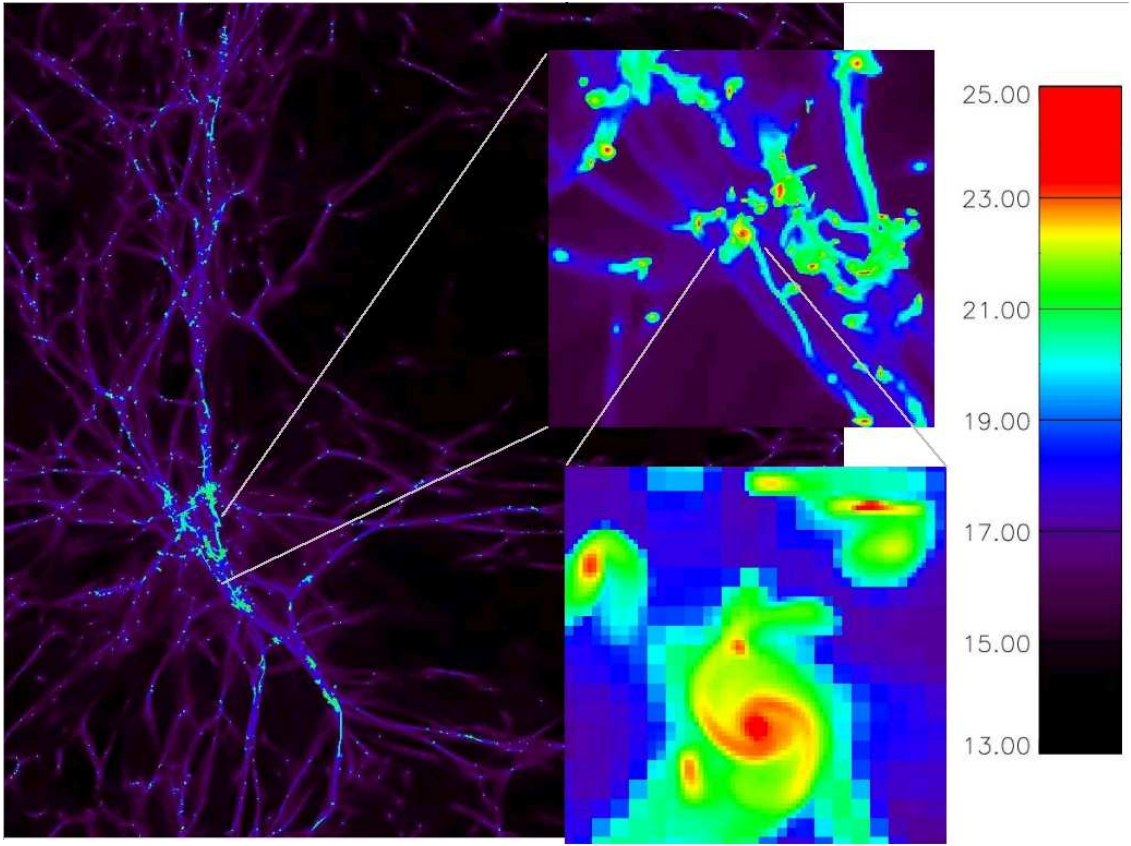}
\caption{HI column density in a volume $8h^{-1}\mpc$ on a side and
  $8h^{-1}\mpc$ thick (comoving), from run C1 at $z=3$. The zoom-in
  boxes are $600h^{-1}\kpc$ (top) and $80h^{-1}\kpc$ (bottom) on a
  side in comoving units, translating into an approximately $10\kpc$
  (physical units) DLA around the disk galaxy in the center of the
  bottom panel. The DLA column densities correspond to green and up in
  the color map on the right. The resolved dynamic range in this model
  is 8192.\label{fig:runC1.z3q10s10.columnDensityMap.mosaic}}
\end{figure*}

To demonstrate the nature of DLAs produced in our models, in
Fig.~\ref{fig:runC1.z3q10s10.columnDensityMap.mosaic} we plotted the
projected HI column density in a volume 8 Mpc on a side and 8 Mpc
thick, for run C1 at $z=3$, with zoom-ins on a cluster of galaxies and
a disk galaxy. All objects with HI column densities above $2\times
10^{20}\cm^{-2}$ give rise to DLAs. To investigate the validity of our
results and compare them to observations, we use two types of
distributions: the HI column density frequency distribution $f(N,X)$
and the line density $\ell_{\rm DLA}(X)$ of DLAs with the Si II
velocity width higher than $v_{\rm Si II}$ vs. $v_{\rm Si II}$. The
frequency distribution $f(N,X)$ is defined such that $f(N,X)dNdX$ is
simply the number of DLAs in the intervals $[N,N+dN]$ and $[X,X+dX]$,
where $dX$ is the ``absorption distance'' interval

\begin{equation}
dX={H_0\over H(z)}(1+z)^2dz.
\end{equation}

In principle, any combination of the following four factors can affect
our results: finite grid resolution, finite mass resolution, the
amplitude and spectrum of the assumed UV background, and radiative and
mechanical feedback from SF. As the goal of this work is to
demonstrate a method to compute the ionization structure of the
outskirts of high redshift galaxies with self-consistent radiative
transfer of the UVB, we do not include the uncertainties of feedback
from SF here. In this paper we concentrate on resolution issues in the
new models and the dependency on the UVB, at a fixed redshift
($z=3$). In our next paper we will study the redshift evolution and
stellar feedback.


\subsection{Numerical resolution}

\begin{figure}
\epsscale{1.}
\plotone{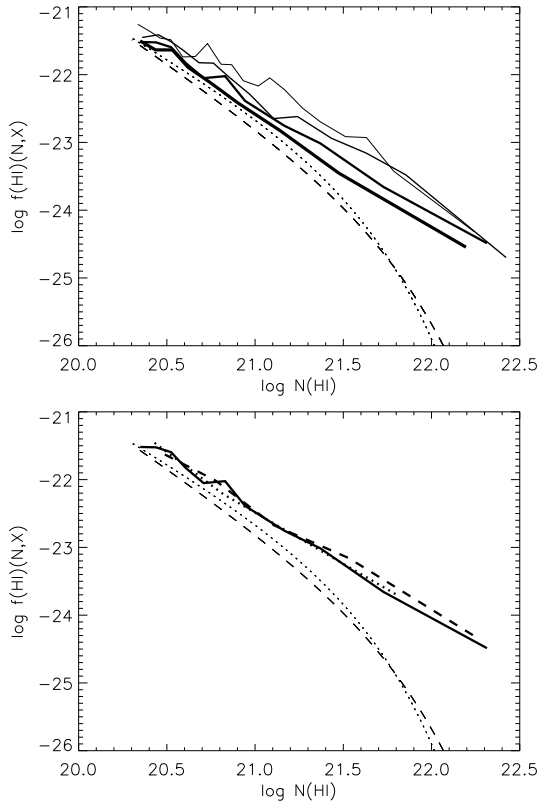}
\caption{Top panel: the upper four solid lines show the HI column
  density frequency distribution $f(N,X)$ for runs C-1 , C0, C1, and
  C2 (from thin to thickest line), all at $z=3$. The lower two lines
  are the $\Gamma$-function fits to the SDSS DLA survey data from
  \citet{prochaska..05} for $z=2.5-3.0$ (dashed) and $z=3.0-3.5$
  (dotted). Lower panel: same as in the top panel, except that the
  upper three lines give $f(N,X)$ for runs C1 (solid), B1 (dashed),
  and A1 (dotted).\label{fig:frequencyDistribution.runC-1.runC0.runC1.runC2.runA1.runB1.z3q10s10}}
\end{figure}

In
Fig.~\ref{fig:frequencyDistribution.runC-1.runC0.runC1.runC2.runA1.runB1.z3q10s10}
we plotted $f(N,X)$ for all of our runs at $z=3$. For comparison, we
also plotted the data from the SDSS DLA survey from
\citet{prochaska..05} for redshift intervals $z=2.5-3.0$ and
$z=3.0-3.5$. The top panel shows our $8h^{-1}\mpc$ models, for which
the grid resolution ranges from $4h^{-1}\kpc$ (comoving) for run C-1
to $0.5h^{-1}\kpc$ for run C2. At this redshift these numbers
translate into 1.5 kpc and 190 pc physical resolution,
respectively. Although the results have not converged yet at our
highest resolution, the trend is clear: as we increase grid
resolution, the baryons in halos can collapse further yielding smaller
DLA cross-sections. For any curve, the disagreement between models and
observations is the largest for high-column density systems which have
the sharpest baryon concentrations and are particularly sensitive to
grid resolution.

The challenging aspect of the DLA modeling is clearly illustrated in
the lower panel in
Fig.~\ref{fig:frequencyDistribution.runC-1.runC0.runC1.runC2.runA1.runB1.z3q10s10}
in which we apply the same $128^3$ base grid with 6 levels of
refinement to a smaller ($4h^{-1}\mpc$) volume, resulting in two times
better grid resolution and eight times higher mass resolution (run B1,
dashed line). Unlike model C1 (solid line), this run has a population
of self-shielded halos in the $10^8-10^9\msun$ mass range. While
naively one would expect to see higher $f(N,X)$, especially at low
column densities, this effect is almost precisely offset by the
reduced size of individual halos through higher grid resolution. This
is further seen in run A1 (dotted line), which extends the population
of self-shielded halos down to $\sim 7\times 10^7\msun$ but features
almost identical $f(N,X)$.

\begin{figure}
\epsscale{1.}
\plotone{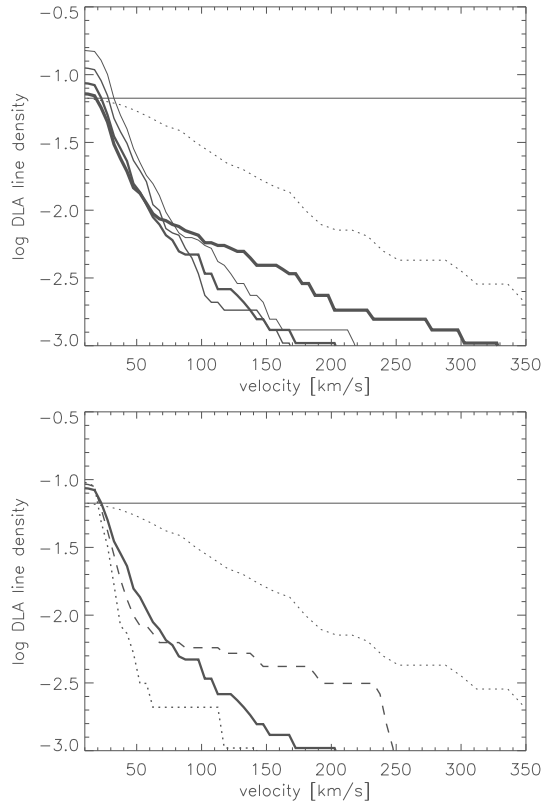}
\caption{Top panel: line density $\ell_{\rm DLA}(X)$ of DLAs with the
  Si II velocity width higher than $v_{\rm Si II}$ vs. $v_{\rm Si
  II}$, at $z=3$. The lines correspond to models C-1 , C0, C1, and C2
  (from thin to thickest line). The dotted line shows the observed
  kinematic distribution from data at all redshifts, compiled from
  Fig. 10 in \citet{wolfe..05} and normalized to match the cumulative
  observed value $0.067\pm 0.006$ at $z=3$ (horizontal line). Lower
  panel: same as in the top panel, except that the three thick lines
  give $\ell_{\rm DLA}(X)$ for runs C1 (solid), B1 (dashed), and A1
  (dotted).\label{fig:abundance.velocity.runC-1.runC0.runC1.runC2.runA1.runB1.z3q10s10}}
\end{figure}

The same effects can be observed in
Fig.~\ref{fig:abundance.velocity.runC-1.runC0.runC1.runC2.runA1.runB1.z3q10s10}
which shows the line density $\ell_{\rm DLA}(X)$ of DLAs per unit
absorption distance with the Si II velocity width higher than $v_{\rm
Si II}$. This figure is similar to the cumulative abundance plots
vs. halo circular velocity (or mass) in \citet{gardner...01} and
\citet{nagamine..04a}, but we chose the neutral gas line width as a
more generic measure of the strength of the absorber, since our
simulated absorbers can also be caused by tidal tails and
filaments. The horizontal lines in
Fig.~\ref{fig:abundance.velocity.runC-1.runC0.runC1.runC2.runA1.runB1.z3q10s10}
show the mean observed DLA line density from \citet{prochaska..05}.

In the top panel we see a monotonic decrease in the line density of
DLAs as we increase the grid resolution. The higher velocity tail is
not affected until we compute our highest resolution model C2 which
starts to resolve individual minihalos in the highest density
regions. With a certain probability that is related to the physical
size of their neutral cores, these small galaxies (with a few tidal
streams mixed in) can be seen as multiple absorption components in a
single line of sight to a remote quasar.

\begin{figure}
\epsscale{.90}
\plotone{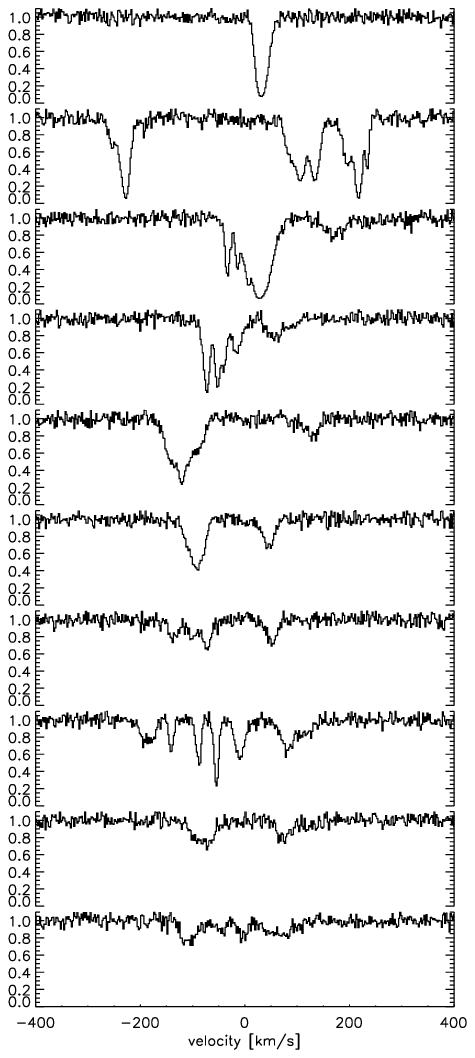}
\caption{Si II $1526\AA$ (or $1808\AA$) line profiles for ten
  selected DLAs in run C2 at $z=3$.\label{fig:ten.runC2.z3q10s10}}
\end{figure}

This effect can be further illustrated in
Fig.~\ref{fig:ten.runC2.z3q10s10} in which we plotted Si II $1526\AA$
or $1808\AA$ line profiles for a typical DLA (top profile) and for
nine DLAs with largest velocity widths (profiles 2-10, from top to
bottom) in model C2. With the line detection criteria defined in
Sec.~\ref{sec:spectrum}, we found the following velocity widths for
these spectra: 31, 468, 191, 149, 281, 168, 170, 306, 193, and 205
$\kms$ (from top to bottom panels in
Fig.~\ref{fig:ten.runC2.z3q10s10}). Spectra 2 and 8 (counting from the
top) are particularly clear examples of several components falling
onto the same line of sight. These multiple component DLAs give rise
to distinct tails at higher velocities which stand out in our high
(190 pc physical) resolution models C2 (upper panel) and B1 (lower
panel). While the highest resolution run A1 (lower panel) shows a
similar tail, it is shifted towards lower ($65-100\kms$) velocity
widths, in part due to the smaller velocity dispersion in the cluster,
and in part due to the lower cross-sections of individual absorbers.

To summarize, none of our models at this stage are able to reproduce
the observed velocity width distribution for systems with $v_{\rm Si
II}\gsim 30\kms$. Although the resolution effects are complex, and
none of our models have fully converged to the observed column density
distribution yet, it seems plausible that some of the effects
described in this section can drive the line profile distribution to
even lower velocities. For example, as we increase grid resolution,
fewer and fewer systems will be crossed by the same line of sight,
moving many DLAs in
Fig.~\ref{fig:abundance.velocity.runC-1.runC0.runC1.runC2.runA1.runB1.z3q10s10}
from $100-400\kms$ into the $30-100\kms$ range. We therefore conclude
that feedback from star formation seems to be the most likely
mechanism to get DLA velocity widths as high as indicated by
observations. In our future simulations, in addition to exploring
feedback, we will also strive to increase mass resolution in
progressively larger simulation volumes, which should produce many
more self-shielded halos with masses of few $10^8\msun$ in galaxy
clusters with larger velocity dispersions.


\subsection{Halo and intergalactic DLAs}

\begin{figure}
\epsscale{1.1}
\plotone{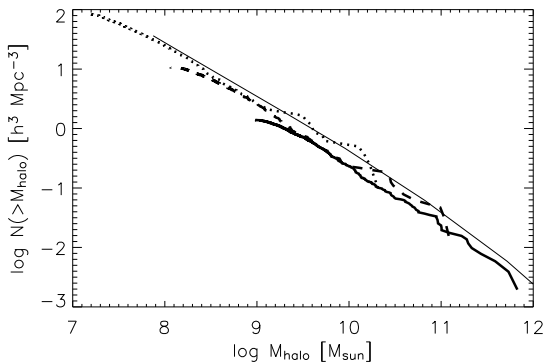}
\caption{Cumulative halo mass function at $z=3$ for dark matter only,
  for runs C1 (thick solid line), B1 (thick dashed line), and A1
  (thick dotted line). The thin solid line is the analytical mass
  function from
  \citet{mo.02}.\label{fig:haloMassFunction.runA1.runB1.runC1}}
\end{figure}

We find that the nature of DLA absorption is a function of the halo
environment. The cumulative halo mass function in our $8h^{-1}$,
$4h^{-1}$ and $2h^{-1}\mpc$ volumes at $z=3$ is plotted in
Fig.~\ref{fig:haloMassFunction.runA1.runB1.runC1}. Most of these halos
give rise to DLAs; however, not all DLAs can be associated with
individual halos.

\citet{gardner...97a} assumed a relation between the neutral hydrogen
cross-section and the host halo mass to predict the total abundance of
DLAs in a large simulation volume. On the other hand,
\citet{nagamine..04a} computed this relation for individual halos in a
series of SPH models with the assumption that all DLAs are fully
neutral and therefore a self-shielding correction is not necessary.

\begin{figure}
\epsscale{1.1}
\plotone{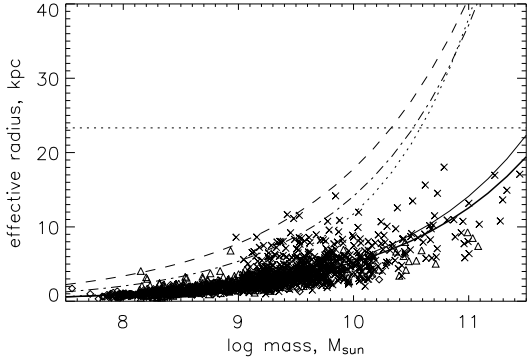}
\caption{Effective radius vs. halo mass, for all DLAs with identified
  host halos in runs A1 (diamonds), B1 (triangles), and C2 (crosses)
  at $z=3$, for the fiducial background in Fig.~\ref{fig:uvb02}. This
  figure does not account for DLAs caused by intergalactic HI clouds
  for which no host halo in the same base grid cell was found. The
  dotted horizontal line shows the size of the base grid cell in the
  $8h^{-1}\mpc$ volume. The solid lines are our least-squares fits
  with eq.~\ref{eq:massCrossSectionFit} for all halos from runs A1, B1
  and C2 (thick line), and all halos from runs A1, B1 and C1 (thin
  line). For comparison, we also plotted the fits to DLA radii from
  \citet{nagamine..04a}, for models O3 (dashed line, no wind, 10 Mpc
  volume), Q3 (dashed-dotted line, strong wind, 10 Mpc volume, low
  resolution) and Q5 (dotted line, strong wind, 10 Mpc volume, high
  resolution).\label{fig:massCrossSection.runA1.runB1.runC2.z3q10s10}}
\end{figure}

Since we include the effects of the UVB on gas at the surface of DLAs,
it is interesting to make an independent estimate of neutral hydrogen
cross-sections for individual absorbers in our models. We draw a large
number ($N_{\rm LOS}=10^8$) of random lines of sight parallel to all
three major axes, and find all absorption systems with the neutral
hydrogen column density above $2\times 10^{20}\cm^{-2}$. We then look
for all halos inside the same base grid cell as the geometrical center
of a DLA, find the one which is closest to this DLA, and associate it
with this absorption system. Some extended DLAs caused by
intergalactic HI will have no host halos. Halos were identified with
the HOP algorithm \citep{eisenstein.98} using the routine
\emph{enzohop} which is part of the \emph{Enzo} package. Following
this procedure, we count the number of damped systems $N_{\rm DLA}$
associated with each halo. Then the effective radius of the absorber
can be estimated approximately as

\begin{equation}
r_{\rm eff}=\sigma^{1/2}\approx L_{\rm box}\left(N_{\rm DLA}\over
N_{\rm LOS}\right)^{1/2},
\end{equation}

\noindent
where $L_{\rm box}$ is the physical box size. In
Fig.~\ref{fig:massCrossSection.runA1.runB1.runC2.z3q10s10} we plotted
a relation between the halo mass $M_{\rm halo}$ and its absorption
radius $r_{\rm eff}$ in runs A1 (diamonds), B1 (triangles), and C2
(crosses). A least-squares fit to these data with equation

\begin{equation}
\log(r_{\rm eff}/\kpc)=\alpha\log(M_{\rm halo}/\msun)+\beta
\label{eq:massCrossSectionFit}
\end{equation}

\noindent
gives $\alpha = 0.38$ and $\beta = -3.07$ (thick solid line). At $z=3$
our absorption cross-sections are approximately a factor of two lower
than the ones in \citet{nagamine..04a}. To avoid any confusion, we
want to stress that this finding does not contradict our DLA counts in
Fig.~\ref{fig:frequencyDistribution.runC-1.runC0.runC1.runC2.runA1.runB1.z3q10s10}
and
\ref{fig:abundance.velocity.runC-1.runC0.runC1.runC2.runA1.runB1.z3q10s10}
since the halos in
Fig.~\ref{fig:massCrossSection.runA1.runB1.runC2.z3q10s10} do not
represent all DLAs in the volume as this plot does not include
intergalactic HI clouds. In addition, our data in
Fig.~\ref{fig:massCrossSection.runA1.runB1.runC2.z3q10s10} should be
viewed as lower limits to DLA cross-sections rather than their actual
values since in a massive cluster environment very often we cannot
identify a single halo responsible for a particular DLA. For example,
in Fig.~\ref{fig:runC1.z3q10s10.columnDensityMap.mosaic} we can see
larger HI clouds engulfing multiple halos. For this reason, in our
estimate of $r_{\rm eff}$ we only searched for halos inside the same
base grid cell as the geometrical center of a DLA which undoubtedly
underestimates the true absorption cross-section.

In addition, interactions between galaxies in more massive
environments create tidal streams which can produce DLA lines. The
cluster in our $8h^{-1}\mpc$ volume at $z=3$ has extended HI tails
formed by galaxy-galaxy interactions which -- at certain line of sight
orientations -- produce column densities in the range $2\times
10^{20}-5\times 10^{21}\cm^{-2}$. The typical physical density in the
self-shielded filaments in these streams is of order $\sim
10^{-2}\cm^{-3}$, the hydrogen neutral fraction often (but not always)
exceeds $0.9$, and physical widths are of order $5-10\kpc$. Gas which
is not self-shielded from the UVB will never reach these column
densities, unless perhaps it is highly compressed into shocks by
feedback which we do not include here.

To estimate the fraction of the line density $\ell_{\rm DLA}(X)$ due
to gas outside of halos, we removed all neutral hydrogen from inside
the virial radii of all halos in the highest resolution $8h^{-1}$ run
C2, and reran the analysis. We defined the virial radius as the radius
of the sphere that encloses 180 times the mean mass density of the
Universe at that redshift. We found that at our highest numerical
resolution approximately 29\% of all DLAs are due to gas outside of
galaxies at $z=3$. While in the original run C2 the highest DLA column
density is $10^{23.8}\cm^{-2}$, it changes to $10^{21.5}\cm^{-2}$ if
we consider only intergalactic gas.

On the other hand, in less massive halo environments, especially in
the smaller $4h^{-1}$ and $2h^{-1}\mpc$ simulation volumes, the main
reason behind our fairly large DLA counts despite the small effective
cross-sections is a much larger contribution to $f(N,X)$ from lower
mass halos in the range $\sim 7\times 10^7-10^8\msun$ which we
demonstrate below.

\subsection{Photoionization of low-mass galaxies}

It is well known that low-mass galaxies cannot retain gas after
reionization. The exact value of the cut-off is not very well
established and without doubt depends on the local
environment. \citet{nagamine..04a} notice a sharp drop-off in the DLA
cross-section below the mass $\sim 10^8 M_\odot$ which they associate
with photoevaporation of gas in these halos by the ionizing UVB and/or
with supernovae feedback.

Among all of our models, the least massive halo which is associated
with a DLA has a dark matter mass $3.6\times 10^7\msun$
(Fig.~\ref{fig:massCrossSection.runA1.runB1.runC2.z3q10s10}). It was
found in run A1 in which the halo mass function extends down to
$1.4\times 10^7\msun$
(Fig.~\ref{fig:haloMassFunction.runA1.runB1.runC1}) which is $\sim 35$
times the mass resolution of this run. The two other least massive
halos associated with DLAs were also found in this highest resolution
run. However, it is only above the mass $\sim 7\times 10^7\msun$ that
we start seeing a significant population of halos with HI in
absorption.

The fact that we see neutral gas in halos with masses slightly below
the cut-off of \citet{nagamine..04a} can be attributed to both the
spatial variations in the UVB inside the galaxy cluster and the lack
of supernova feedback in our models. As \citet{nagamine..04a} pointed
out, a sharp cut-off should be expected due to the strong dependence
of baryon cooling and photoheating on the virial temperatures of halos
around $\sim 10^4{\rm K}$. In fact, we see such a cutoff at $\sim
7\times 10^7\msun$ in
Fig.~\ref{fig:massCrossSection.runA1.runB1.runC2.z3q10s10}. As all
photoionizing radiation in our models comes from the box boundary, few
low-mass halos might be shadowed by other more massive systems and be
able to retain their neutral gas, as we see in two or three galaxies
with masses below $7\times 10^7\msun$. Furthermore, a number of
galaxies in the mass range $7\times 10^7\msun-10^8\msun$ are not
exposed to the full UVB as part of their ``sky'' is blocked by nearby
absorption systems (see, e.g., the zoom-in panels in
Fig.~\ref{fig:runC1.z3q10s10.columnDensityMap.mosaic}), and the mass
cut-off above which galaxies can retain neutral gas is shifted towards
slightly lower masses. Of course, this effect depends on the
environment and will be more pronounced at higher redshifts where the
spatial variations in the UVB are larger. Star formation and supernova
feedback will partially reverse this effect removing neutral material
from lower mass galaxies assuming that these galaxies have accumulated
enough cold gas to host star formation in the first place. However,
this effect is very complex and, in addition to the mechanical
feedback from winds and supernovae, will have to include transfer of
locally generated UV photons.

\subsection{Dependence on the UVB}

\begin{figure}
\epsscale{1.}
\plotone{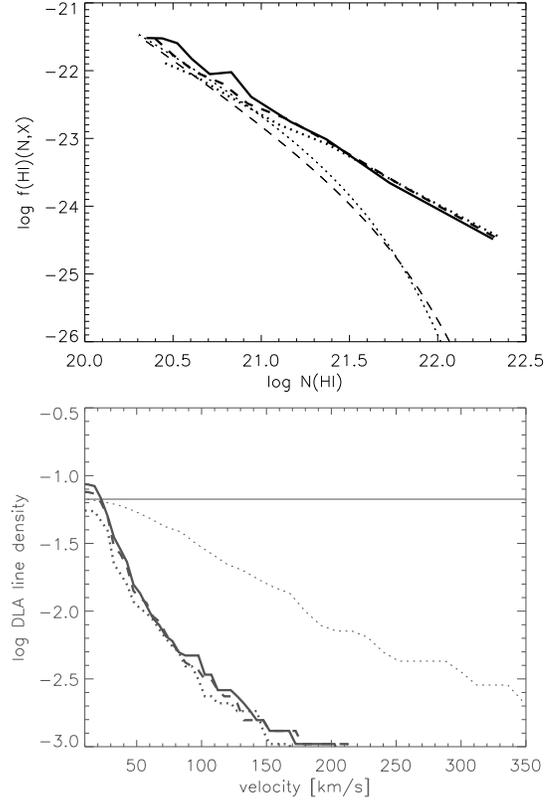}
\caption{The three thick lines in each panel show results for a run
  with the full quasar background from Fig.~\ref{fig:uvb02} (C1, solid
  line), and runs with two times (C1b, dashed line) and four times
  (C1c, dotted line) the fiducial quasar background, all at $z=3$. Top
  panel: HI column density frequency distribution $f(N,X)$. The two
  thin lines are the SDSS DLA survey data as in
  Fig.~\ref{fig:frequencyDistribution.runC-1.runC0.runC1.runC2.runA1.runB1.z3q10s10}.
  Lower panel: line density $\ell_{\rm DLA}(X)$ of DLAs with the Si II
  velocity width higher than $v_{\rm Si II}$ vs. $v_{\rm Si II}$. The
  thin dotted and thin horizontal lines show the observed kinematic
  distribution as in
  Fig.~\ref{fig:abundance.velocity.runC-1.runC0.runC1.runC2.runA1.runB1.z3q10s10}.
  \label{fig:frequencyDistribution.abundanceVelocity.runC1.z3q10s10.z3q20s10.z3q40s10}}
\end{figure}

Our distributions are somewhat sensitive to a change in the UVB. In
model C1 we varied the quasar component from the fiducial value in
Fig.~\ref{fig:uvb02} to twice (C1b) and four times (C1c) that
amplitude and plotted the results in
Fig.~\ref{fig:frequencyDistribution.abundanceVelocity.runC1.z3q10s10.z3q20s10.z3q40s10}.
As we increase the UVB amplitude and simultaneously steepen its
spectrum, more HI is ionized everywhere on the outskirts of galaxies,
but in terms of DLA statistics -- not surprisingly -- primarily low
column density systems are affected. While the under-resolved model C1
features the cumulative number of DLAs $\sim25\%$ above the observed
value at $z=3$, the higher UVB model C1c has the number of DLAs
$\sim25\%$ below the observed value. Noticeable differences between
the two models extend to the column density $10^{21.4}\cm^{-2}$, but
the higher column density systems are not affected.

\subsection{Effects of radiative transfer}

\begin{figure}
\epsscale{1.}
\plotone{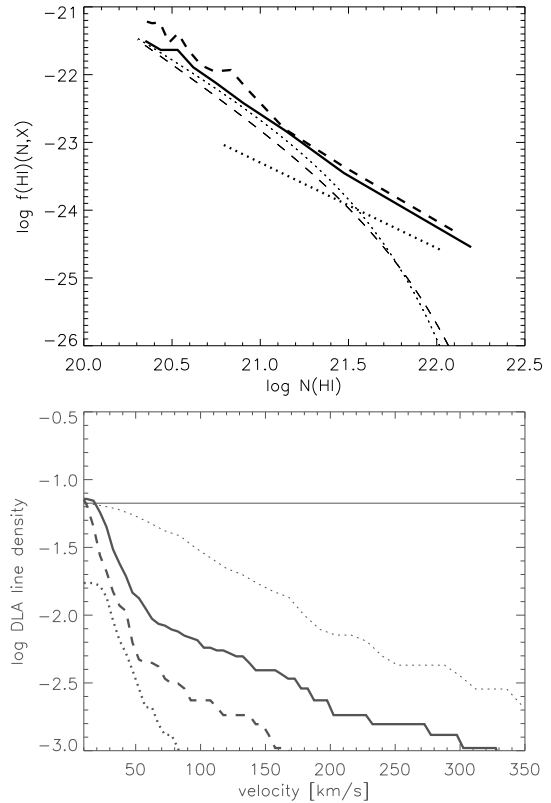}
\caption{The three thick lines in each panel show results for a run
  with full radiative transfer (C2, solid line), a run with
  self-shielding above the physical density $10^{-2}\cm^{-3}$ (C2s,
  dashed line), and a run with the optically thin approximation
  throughout the entire simulation volume (C2t, dotted line), all at
  $z=3$. Top panel: HI column density frequency distribution
  $f(N,X)$. The two thin lines are the SDSS DLA survey data as in
  Fig.~\ref{fig:frequencyDistribution.runC-1.runC0.runC1.runC2.runA1.runB1.z3q10s10}.
  Lower panel: line density $\ell_{\rm DLA}(X)$ of DLAs with the Si II
  velocity width higher than $v_{\rm Si II}$ vs. $v_{\rm Si II}$. The
  thin dotted and thin horizontal lines show the observed kinematic
  distribution as in
  Fig.~\ref{fig:abundance.velocity.runC-1.runC0.runC1.runC2.runA1.runB1.z3q10s10}.
  \label{fig:frequencyDistribution.abundanceVelocity.runC2.z3q10s10.shield.thin}}
\end{figure}

In addition to the full radiative transfer run C2 in the $8h^{-1}\mpc$
volume, we also computed a model C2s with complete self-shielding
above the physical density $10^{-2}\cm^{-3}$, assuming that the local
UVB is just the uniform cosmic background from Fig.~\ref{fig:uvb02} in
every cell below this threshold, and zero above it. This assumption
has been previously used in DLA modeling, e.g., in
\citet{haehnelt..98}. We also computed a model C2t with the optically
thin approximation imposing the same uniform cosmic UVB in every cell
in the volume. In
Fig.~\ref{fig:frequencyDistribution.abundanceVelocity.runC2.z3q10s10.shield.thin}
we plotted distributions for these three models. All three models were
iterated until H/He ionization equilibrium.

It is evident that self-shielding of neutral gas is the dominant
mechanism determining all properties of DLAs. The model with the
uniform background in the optically thin regime produces very few DLAs
as expected. Surprisingly, the full transfer model and the one with
self-shielding give roughly the same cumulative number of
DLAs. However, since the transfer model includes high energy photons,
the ionizational structure of HI regions is more complex than in the
fully shielded case. These differences result in vastly different line
widths distributions: in the self-shielded model most lines widths are
clustered around $15\km\s^{-1}$, whereas in the model with transfer
they are concentrated at higher velocities in the range
$15-30\kms$. It is reassuring that the model with full transfer in the
lower panel in
Fig.~\ref{fig:frequencyDistribution.abundanceVelocity.runC2.z3q10s10.shield.thin}
is about half-way from the self-shielded model to observational data.

It is important to point out that our self-shielded model produces
results very different from \citet{haehnelt..98}, namely substantially
smaller velocity widths on average. A number of factors could
contribute to this difference, such as the use of Eulerian
hydrodynamics with AMR instead of SPH, but perhaps the most important
factors are our much larger sample of galaxies (we used 710 distinct
halos to generate 406 DLAs in runC2s vs. their sample of 40
protogalactic clumps used repeatedly to produce 640 DLAs), and our
much higher grid resolution (190 pc physical vs. their 1 kpc). But as
we point out, our full radiative transfer runs produce further
improvement in kinematics modeling rendering any self-shielded models
obsolete.

\begin{figure}
\epsscale{1.}
\plotone{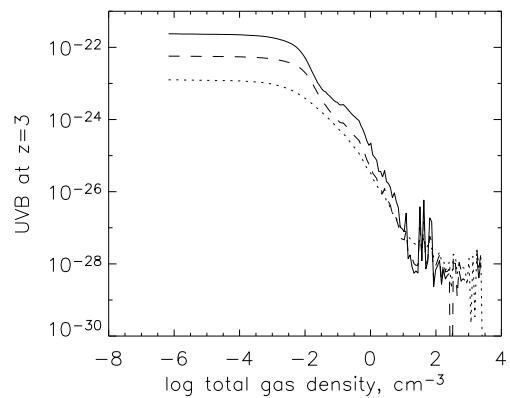}
\caption{Mean specific intensity
  ($\erg\s^{-1}\cm^{-2}\hz^{-1}\sr^{-1}$) vs. gas number density in
  model C1 at $z=3$ at photon energies $13.6\ev$ (solid line),
  $24.6\ev$ (dashed line) and $54.4\ev$ (dotted line).\label{fig:jmean.runC2.z3q10s10}}
\end{figure}

In Fig.~\ref{fig:jmean.runC2.z3q10s10} we plotted the mean specific
intensity at $13.6\ev$, $24.6\ev$ and $54.4\ev$ vs. the local gas
number density at $z=3$. One can easily see hardening of the spectrum
in self-shielded regions above the physical density $10^{-2}\cm^{-3}$,
as more energetic photons travel further in the neutral medium.

\section{Summary}\label{Summary}

The standard hierarchical model is very successful in explaining the
emergence of structures in the early Universe. DLAs are viewed as
neutral gas clouds confined to individual galaxies, whether in
virialized systems or systems still in the process of merging. This
paradigm is clearly seen in most previous DLA simulation literature
where analysis is based on a fit to the relation between DLA
cross-section and halo mass. The results of high resolution
simulations by \citet{haehnelt..98} suggest that small protogalactic
clumps in the process of merger could explain the observed line
profiles, and the statistical analysis of absorption line kinematics
within semi-analytical models by \citet{maller...01} suggests that
tidal tails can be responsible for absorption.


To test both these hypotheses and a broader idea that DLAs can be
caused by absorption in neutral intergalactic clouds either stripped
of galaxies during interactions or even still accreting onto galaxies
from the IGM, we propose to include radiative transfer of the UVB into
galaxy formation models, and argue that this new piece of physics is
essential in order to reproduce DLA observables such as the column
density and velocity width distributions. An accurate treatment of
background photons allows us to address the ionization structure of
the outskirts of high redshift galaxies in a much more precise way.

We simulate galaxy formation in a series of cosmological volumes
ranging in size from 2 Mpc to 8 Mpc, solving simultaneously for the
first time coupled equations of hydrodynamics, self-gravity,
multi-species chemistry and radiative transfer to account for
shielding against the UVB. Radiative transfer is computed on all
levels of grid refinement with a two-stage calculation, first just
along the three major axes simultaneously with hydrodynamics, and then
in 192 directions in the postprocessing mode. We use AMR to resolve
individual galaxies with the same refinement criteria everywhere in
the volume. Similar to \citet{nagamine..04a}, we use a series of
computational volumes of different sizes while varying numerical
resolution to try to achieve optimal balance between resolution and
fair volume sampling. However, because we do not restrict damped
Ly$\alpha$ absorption to halos, we do not assume any single relation
between DLA cross section and halo mass for analysis. There is a
relation between DLA cross-section and associated line width which in
{\it some} cases is indeed caused by the velocity dispersion within
individual galaxies. Our findings are as follows.

\begin{itemize}

\item

  {\bf Resolution:} As we increase grid resolution, our column
  densities demonstrate a trend in convergence towards the observed
  distributions. However, we cannot reproduce the high-end tail of the
  velocity width distribution. Although it is possible that the
  resolution effects and limitation in box sizes are responsible for
  this shortcoming, feedback from star formation which we have not
  included into the current models is likely to produce expanding
  shells and therefore higher velocity widths on average. The column
  density distributions alone indicate that the minimum required grid
  resolution has to be better than $\sim 500\pc$ in comoving
  coordinates, or $\sim 100\pc$ in physical space. The minimum mass
  resolution is $\sim 10^6\msun$ corresponding to at least a hundred
  DM particles per halo which is still capable of being self-shielded
  from the UVB.

\item
  {\bf Filamentary DLAs:} The nature of DLA absorption is a function
  of halo environment, at any given redshift. Although all
  self-shielded halos can give rise to DLAs, not all DLAs can be
  associated with individual halos. We find that in massive cluster
  environments at $z=3$ tidal tails and quasi-filamentary structures
  often appear to be self-shielded against the UVB. The typical column
  densities of filamentary DLAs are below $10^{21}\cm^{-2}$, and
  physical widths are of order $5-10\kpc$. Since we have not achieved
  numerical convergence, we have not attempted a comprehensive,
  redshift-dependent statistical study to distinguish halo DLAs from
  the ones caused by tidal tails, and both of these from DLAs caused
  by primordial gas still accreting onto galaxies.

  Provided that filamentary structures are still self-shielded at
  higher grid resolution, it seems plausible that going to higher
  resolution would only increase the filamentary contribution to DLA
  absorption, as the 3D density peaks (halos) would contribute
  progressively less to the total HI column density than the 2D
  density peaks (filaments) as baryons cool and collapse on finer
  grids.

\item
  {\bf Photoionization of low-mass galaxies:} The least massive single
  galaxy associated with DLAs in our models has the mass $3.6\times
  10^7\msun$. However, we see a significant population of halos with
  HI in absorption only above the mass $\sim 7\times 10^7\msun$ below
  which just a few dwarf galaxies retained any sufficient amount of
  neutral gas by $z=3$. This threshold is slightly lower than the
  previously reported values of $10^8\msun$ to few $10^8\msun$
  \citep{nagamine..04a} contributing to larger DLA counts in less
  massive halo environments. Part of this discrepancy can be explained
  by spatial variations in the UVB inside the galaxy cluster, and it
  is possible that supernova feedback which is not included into our
  models could also have a role removing neutral gas from low-mass
  galaxies.

\item
  {\bf Self-shielding model:} Calculations with self-shielding above
  density $10^{-2}\cm^{-3}$ produce roughly the same total number of
  DLAs per unit absorption distance as the run with full radiative
  transfer. However, the absorption line width distributions in the
  two models are markedly different due to variations in ionizational
  profiles.

\end{itemize}

We intend this work to be the first in a series of papers exploring
DLA properties with our new cosmological simulations with high-angular
resolution radiative transfer on adaptively refined meshes. A number
of topics are open to further exploration.

1) At the moment it is not clear to what extent DLAs produced with
full coupled RHD simulations of galaxy formation would be different
from the ones we get with our radiative-chemical postprocessing
approach. Inclusion of radiative terms into hydrodynamical equations
might be important for both accessing the effect of the external UV
irradiation of primordial galaxies \citep{iliev..05} and studying
feedback from SF \citep{whalen..04,oshea...05}.

2) We have not included SF into our current models. Its mechanical and
radiative feedback could change the ionization structure and
kinematics of gas in high-redshift galaxies. For example, winds from
SF regions and/or SNe could create denser shells around low-density
bubbles. If dense enough, these shells might contribute to an overall
increase in DLA effective cross-sections \citep{schaye01}. Of course,
there is a competing effect if feedback destroys neutral gas in halos
as observed in \citet{nagamine..04a}.

3) Besides an obvious merit of providing more statistics, going to
larger simulation volumes of few tens $\mpc$ would allow us to test
the effect of large-scale fluctuations in the UVB on absorption
properties of high-redshift galaxies. The average separation between
quasars at $z=3-4$ is more than an order of magnitude greater than our
current largest simulation volume. Similar to the proximity effect
observed in quasar spectra, very few halos would produce a DLA in a
cluster that hosts a quasar. As we move further away from a quasar,
the mean background drops. It is also attenuated in large clusters of
galaxies due to the overall higher IGM column densities. If the
typical UVB in the vicinity of DLAs is 50-75\% lower than the cosmic
average, it could yield dramatically different DLA counts. In
addition, larger simulation volumes would produce more massive
clusters of galaxies, with more violent interactions and more
prominent tidal tails which would also have a direct impact on DLA
observables.

4) To restrict our parameter space, we omitted an interesting topic of
fine-tuning the initial power spectrum (particularly on galactic
scales) to match the observed line density of DLAs which we are
planning to do in the future.

In order to derive a single realistic population of hydrogen absorbers
from Ly$\alpha$ forest clouds to DLAs, ideally one would like to build
a sub-kpc resolution model with both stellar feedback and UVB
radiative transfer in a large (50-80 Mpc) volume. To achieve the
required mass and grid resolution, one would need to use grids of
sizes 2048L6 with $2048^3$ DM particles. An N-body model of this size
(but in a much larger volume) has been recently computed by the Virgo
Consortium -- the ``Millennium Simulation''
\citep{springel................05} -- and can be potentially
postprocessed with our fully threaded transport engine following the
routine utilized in this paper.

\section{Acknowledgments}

This work was supported by Scientific Discovery Through Advanced
Computing (SciDAC), a program of the Office of Science of the U.S.
Department of Energy (DoE); and by Oak Ridge National Laboratory,
managed by UT-Battelle, LLC, for the DoE under contract
DE-AC05-00OR22725. At the beginning of this project AOR was partially
supported by NSF grants AST-9803137 and AST-0307690. JXP and AMW are
partially supported by NSF grant AST-03-07824. Numerical simulations
were performed using the IBM DataStar system at the San Diego
Supercomputer Center.

\bibliographystyle{apj}                       

\end{document}